# Direct observation of electron capture & emission processes by the time domain charge pumping measurement of $MoS_2$ FET

Koki Taniguchi,[1] Nan Fang[1] and Kosuke Nagashio[1,*]
[1]Department of Materials Engineering, The University of Tokyo, Tokyo 113-8656, Japan
**E-mail:** *nagashio@material.t.u-tokyo.ac.jp

**Abstract** Understanding interface properties in $MoS_2$ field effect transistors with a high-*k* gate insulator is critical for improving the performance of the device. Here, by applying the time domain charge pumping method, the elementary process for capture and emission of electrons to the interface states is monitored directly using a fast acquisition system. The main outcome is the detection of the clear difference in the capture and emission process of electrons to the interface states. In addition to the transient current response for gate capacitance, the current peak is observed during electron capture, while the broad tail is detected during electron emission. This different behavior is associated with the fact that the time constant for electron capture is much shorter than that for electron emission. Moreover, $D_{it}$ is evaluated to be in the range of $10^{12}$ - $10^{13}$ $cm^{-2}$ $eV^{-1}$, which is comparable with that estimated from subthreshold swing.

The demonstration of a natural thin-body $MoS_2$ field-effect transistor (FET) with an effective channel length of ~3.9 nm has facilitated research on 2-dimensional (2D) layered channels due to overcoming the scaling limit of ~5 nm for Si gate length.[1] The dominant issue necessary for practical application is improvement in the high-*k*/$MoS_2$ interface properties, even though there are many other issues to be solved due to the nascence of 2D research. Ideally, the dangling bond-free surface of the layered channel is expected to provide an electrically inert interface. However, by using capacitance-voltage (*C-V*) measurements in the multilayer $MoS_2$ metal-oxide-semiconductor capacitor structure,[2-5] the frequency dispersion of the top gate capacitance is observed. The interface trap densities ($D_{it}$) in the range of ~$10^{12}$ - $10^{13}$ $cm^{-2}$ $eV^{-1}$ have been extracted and show band tail energy dispersion and often an additional hump. On the other hand, the recent *C-V* study on $MoS_2$ FET structure[6] has revealed that the frequency dispersion observed in previous literature[7-9] resulted from the high channel resistance, thereby making it difficult to extract $D_{it}$ in a straightforward manner. Based on these trial-and-error efforts, several physical origins, such as bond bending of the Mo *d* orbital,[9] S vacancy,[3,5,10] and trap sites in the oxide, have been proposed. To further understand the origin for the interface states, it is important to analyze the elementary process for capture and emission of electrons, which is obscured in the conventional *C-V* measurement because the time-averaged steady-state current is measured.

Charge pumping (CP) is one of the most widely used techniques for analyzing the interface states of metal-oxide-semiconductor FETs (MOSFET).[11-13] In CP, during the single gate voltage ($V_G$) pulse between accumulation and inversion, electrons that flow into the conduction band are captured in interface states, and some of these electrons recombine with holes that flow into the valence band, as shown in Supplementary **figure S1**. That is, the current flows from the substrate to the source and drain through the interface states. Experimentally, by applying the $V_G$ pulse string using a pulse generator, this time-averaged recombination current (CP current) is monitored by a direct current ammeter, and the interface defect density ($N_{it}$, $cm^{-2}$) can be evaluated. Recently, CP has been examined with respect to time domain measurement, where the transient currents are monitored directly during a single $V_G$ pulse with ramping and falling times of ~10 μs using a fast acquisition system.[14-16] The main outcome is the detection of the clear difference in the capture and emission process of electrons to the interface states. In addition to the transient current response for gate capacitance, the current peak is observed during electron capture, while the broad tail is detected during electron emission. This different behavior is associated with the fact that the time constant for electron capture is much shorter than that for electron emission.

Here, conventional CP is not applicable to the monolayer $MoS_2$ FET, because the source/drain electrodes are connected only to the conduction band due to Fermi level pinning.[17,18] Therefore, the recombination current cannot be measured. On the other hand, time domain CP is possible since the behavior of electrons that flow in and out through the source and drain can be directly monitored. Although pulse *I-V* measurements, which is one of time resolved measurements, have been applied to Si,[19,20] graphene,[21-23] and $MoS_2$[24,25] so far, time domain CP has not been studied in 2D material systems. In this study, time domain CP is applied to monolayer $MoS_2$ FETs to elucidate the carrier response to the interface trap sites during a single $V_G$ pulse by analyzing the transient response of the top gate capacitor. Moreover, $D_{it}$ is also evaluated to compare the data by different methods.

Monolayer $MoS_2$ films were mechanically exfoliated onto a quartz substrate from natural bulk $MoS_2$ flakes.[6] Although it is slightly difficult to find the monolayer using an optical microscope due to the transparent substrate, it is still possible. Ni/Au were deposited as the source/drain electrodes. For the top gate formation, a $Y_2O_3$ buffer layer was first deposited, and then, an $Al_2O_3$ oxide layer with a thickness of 10 nm was deposited via atomic layer deposition (ALD), followed by the Al top-gate electrode formation. The detailed procedure for the top gate formation can be found elsewhere.[26,27] The device image can be seen in **Figure 1**. Alternatively, the monolayer $MoS_2$ FET is also fabricated on a $SiO_2/n^+$-Si substrate for comparison. The *I*-*V* and *C*-*V* measurements were conducted using Keysight B1500 and 4980A LCR meters, respectively. For time domain CP, a Keysight B1530A WGFMU with minimum time resolution of 10 ns was used. All electrical measurements were performed in a vacuum prober at room temperature, where the measurement cable length from the acquisition system to the probe tip (~650 mm) is kept as short as possible to minimize signal reflection.

First, time domain CP was carried out using the Si N-MOSFET test device with intentional electrical stress to confirm the measurement system. The time domain CP is conducted, where the sharp peak during the electron capture and broad tail during the electron emission are evident (see Supplementary **Figure S2**). This preliminary measurement is consistent with previous results,[14,15] suggesting the validity of the present set up.

Now, let us move to the time domain CP for the monolayer $MoS_2$ FET. **Figure 2(a)** shows the transfer characteristics of $MoS_2$ FET on the quartz substrate, where the typical *n*-channel behavior is observed. The two-terminal field effect mobility ($\mu_{FE}$), subthreshold swing (*S.S.*), and threshold voltage ($V_{TH}$) are ~7 $cm^2 V^{-1} s^{-1}$, ~210 $mVdec^{-1}$, and -2.2 V, respectively. For the $\mu_{FE}$ extraction, the contribution of quantum capacitance is neglected, and $C_{TG}$ is determined to be 0.22 $\mu Fcm^{-2}$ as the capacitance at the accumulation in the *C*-*V* measurement.[9] For time domain CP, a single top gate voltage ($V_{TG}$) pulse is applied, while the current through the source and drain ($I_{S/D}$) is measured using the fast acquisition set up, as shown in **Figure 1(c)**. **Figure 2(b)** schematically illustrates the single $V_{TG}$ pulse and the $I_{S/D}$ as a function of time. The final voltage ($V_{fin}$) is always fixed to be 5 V, while the initial voltage ($V_{init}$) is changed in the range from -5 V to 3 V. The area integral of the hatched region ideally indicates the charge ($Q$) accumulated in the $MoS_2$ by the gate capacitor. $Q$ is simply estimated here by $Q = C_{TG} \times \Delta V_{TG}$, where $\Delta V_{TG}$ is defined as $V_{fin} - V_{init}$. **Figure 2(c)** shows $Q$ estimated as a function of $\Delta V_{TG}$ for the $V_{TG}$ ramping case with $t_r = t_w = t_f$ = 10 µs, where $t_r$, $t_w$, and $t_f$ are the ramping time, the waiting time, and the falling time, respectively. The red solid circle shows $Q$ obtained for $MoS_2$ FET on the quartz substrate, which agrees well with $Q$ calculated using $C_{TG}$ = 0.22 $\mu Fcm^{-2}$. On the other hand, the blue rectangles show $Q$ obtained for another $MoS_2$ FET on the $SiO_2/n^+$-Si substrate, which largely deviates from the calculated line.

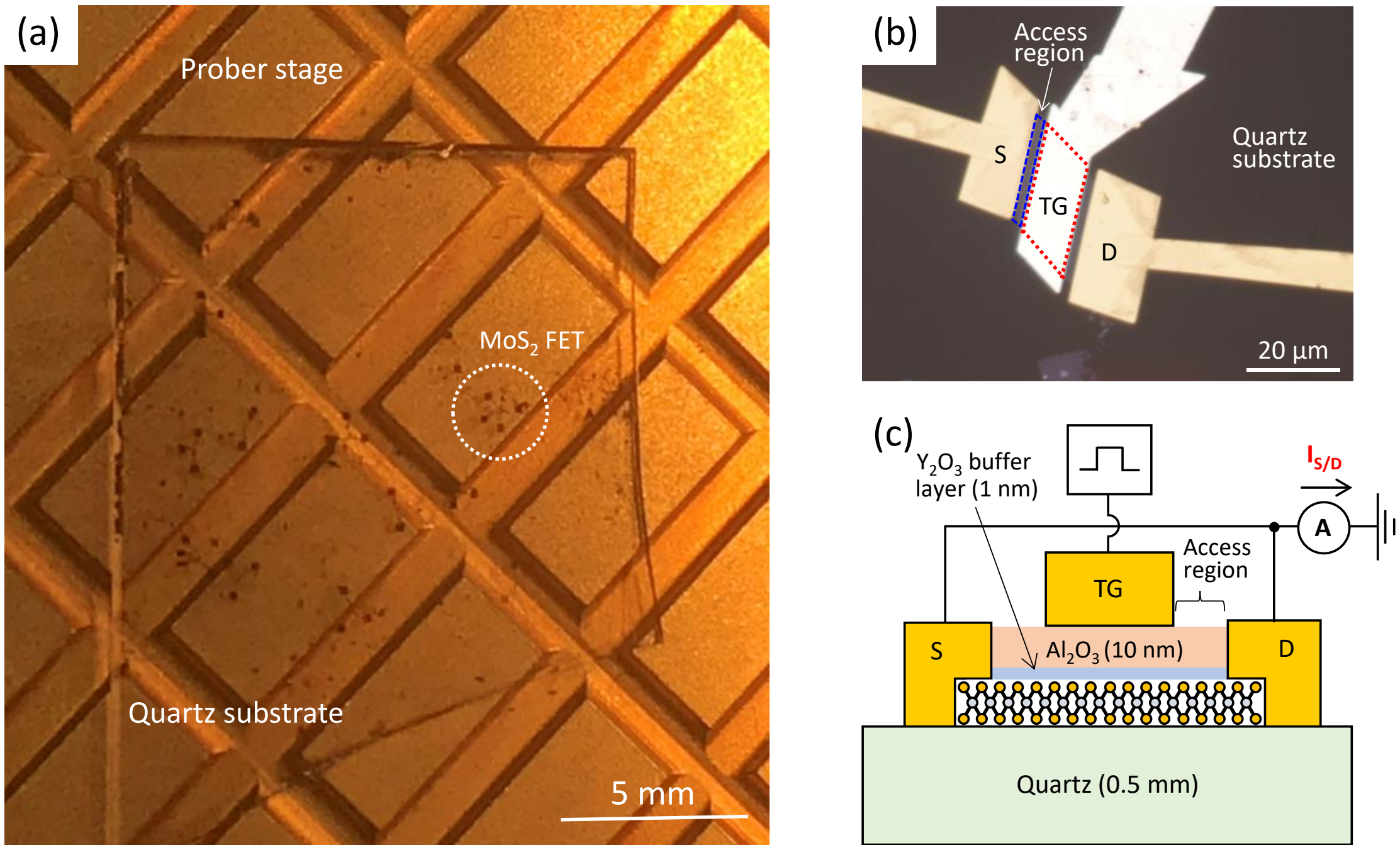

**Figure 1.** (a) Optical micrograph of the device on the quartz substrate. (b) Magnified image of monolayer $MoS_2$ FET with the $Al_2O_3$ top gate insulator. (c) Schematic drawing of $MoS_2$ top gate transistor and measurement set up. The source and drain are tied together, and a pulse voltage is applied to the gate. It should be noted that conventional CP cannot be applied to the present device structure because hole conduction is prohibited by the *n*-type access region as well as the *n*-type contact.

This is due to the large parasitic capacitance ($C_{\mathrm{Para}}$) between the electrode pad and $n^+$-Si substrate. These results suggest that utilization of the highly insulating quartz substrate is critical to extract the correct transient current during high-frequency measurement of a few microseconds.

Since the quantitative validity of the macroscopic transient current for the gate capacitor is confirmed, the elementary processes of electron capture to the interface states and electron emission from the interface states, which are expected in the transient current response, are examined herein. **Figure 3(a)** shows $I_{\mathrm{S/D}}$ as a function of time at different time conditions at the fixed $V_{\mathrm{TG}}$ pulse of $V_{\mathrm{init}}$ = -5 V and $V_{\mathrm{fin}}$ = 5 V. The transient current curves for the falling stage are inverted for comparison with the transient curves for the ramping stage. For $t_{\mathrm{r}} = t_{\mathrm{w}} = t_{\mathrm{f}}$ = 1,000 μs, the signal to noise ratio (S/N) is relatively poor, which slightly blinds the signature on the interaction between electrons and the interface states. It should be noted that the time domain charge pumping method requires large current level to get the high time resolution. Therefore, the large top gate area (> 250 μm$^2$) and short access region length are intentionally prepared, and the cables are connected to both the source and drain electrodes, as shown in **Figures 1(b) and (c)**. Then, the adequately high S/N has been obtained. For $t_{\mathrm{r}} = t_{\mathrm{w}} = t_{\mathrm{f}}$ = 100 μs, a sharp peak in the transient charging current is clearly observed for electron capture as shown by an arrow, while the broad tail behavior is evident as shown by the dotted circle. For a shorter pulse of $t_{\mathrm{r}} = t_{\mathrm{w}} = t_{\mathrm{f}}$ = 10 μs, the current peak becomes stronger. This is because $D_{\mathrm{it}}$ increases close to the conduction band.[9] Finally, it becomes very broad for $t_{\mathrm{r}} = t_{\mathrm{w}} = t_{\mathrm{f}}$ = 1 μs (not shown here) because the measurement current level is insufficient for this short pulse time due to the limited $MoS_2$ channel area. The current peak is observed during the voltage ramping because some of the electrons accumulated in the conduction band are instantly trapped in the interface states, leading to the current peak, as shown schematically in **Figure 2(b)**. On the other hand, the broad current tail during the voltage falling can be explained by the fact that the time constant for electron emission ($\tau_{\mathrm{e}}$) is much larger than that for electron capture ($\tau_{\mathrm{c}}$). It should be noted that this visible band tail is only a small part of the distribution of the electron emission times. This can easily be seen in **Figure 3** for short rise/fall times since the area under both curves differs. Therefore, the major part of the electron emission most likely takes place during $V_{int}$ and is below the measurement resolution. These behaviors are consistent with those observed in Si-MOSFET. Even for the 2D layered channel "ideally" without the dangling bond, the elementary process for interaction between the electrons and the interface states is suggested to be similar to that for Si. It should be noted that the observation of the current peak in Si is possible only after the interface states are intentionally formed by electrical stress.[15,16] Therefore, the present stage of the high-$k$/$MoS_2$ interface quality is worse than that for Si.

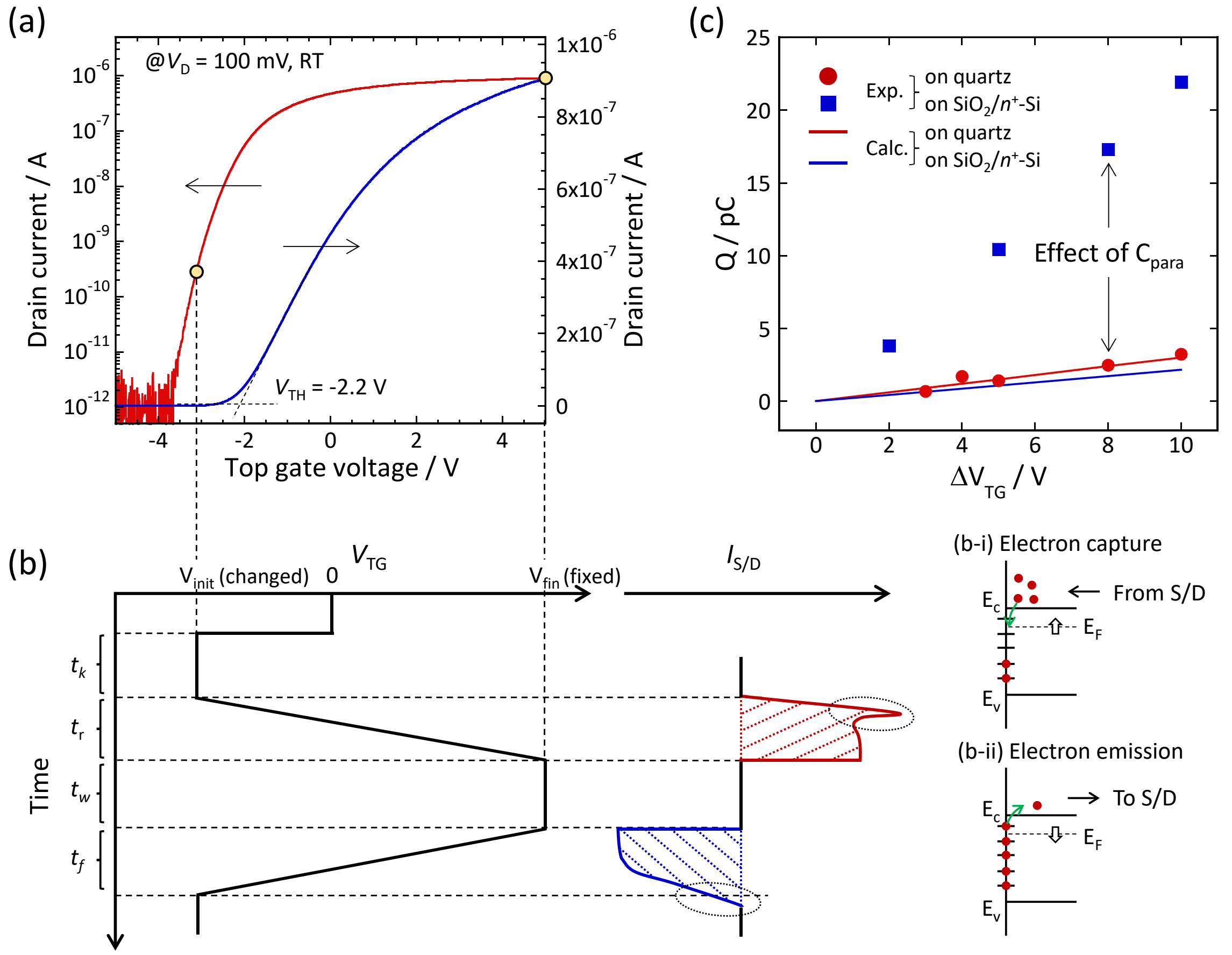

**Figure 2.** (a) Transfer characteristics of monolayer $MoS_2$ FET on the quartz substrate. (b) Schematic illustration for $V_{\mathrm{TG}}$ and $I_{\mathrm{S/D}}$ as a function of time, and for electron capture and emission. Schematic illustration for electron capture and emission behavior is also shown. (c) The charge $Q$ stored for the top gate capacitor as a function of $\Delta V_{\mathrm{TG}}$.

The proper time condition with relatively high S/N is $t_r = t_w = t_f = 100$ µs. Next, under this time condition, $V_{init}$ is altered from depletion to $V_{TH}$ to reveal how the transient current response changes. As shown in **Figure 3(b)**, the current peak gradually disappears at around $V_{init}$ = -3.4 V (not shown here) since the trap sites at the interface have already been occupied at the large initial Fermi energy ($E_F$). At the same time, the broad band tail also disappears because $E_F$ remains above the trap state. $V_{TG}$ at the current peak ($V_{peak}$) is calculated and plotted in **Figure 3(c)**. Although the time at the current peak position shifts for different $V_{init}$, $V_{peak}$ is almost constant in the subthreshold region for all $V_{init}$ values. This is because the ramping time is held constant ($t_r$ = 100 µs). Once $t_r$ (= $t_w = t_f$) is reduced from $10^3$ to $10^1$ µs, $V_{peak}$ shifts closer in value to $V_{TH}$, but remains more negative than $V_{TH}$. This indicates that the electron capture to the interface states finishes before the channel is completely formed. Moreover, this is the origin for the gentle slope of *S.S.*, which is often encountered in transfer curves obtained by DC measurement.[9]

Since the electron capture and emission processes are confirmed in the transient current response, the interface quality should be evaluated quantitatively. In the case of conventional CP,[11,12,16] $N_{it}$ is estimated, which is the interface trap density averaged through the "whole" energy gap. In case of $MoS_2$ FET, however, the recombination current during a gate voltage pulse between the conduction and valence bands cannot be obtained because there is no electrical connection to the valence band. Therefore, let us extract $D_{it}$ as a function of $V_{TG}$. The key feature is that the area integral observed for the transient current during $V_{TG}$ ramping is larger than that observed during falling $V_{TG}$, because most of the broadly distributed electron emission takes place during $V_{int}$ and is therefore below the measurement range. **Figure 4(a)** shows a schematic illustration for $D_{it}$ as a function of $V_{TG}$, where the band tail distribution of $D_{it}$ is adapted for simplicity.[9] In the capture process, all the interface states swept from $V_{init}$ to $V_{fin}$ capture electrons, as shown by the region (i). On the other hand, a deeper energy of the interface states in the emission process leads to a longer $\tau_e$. Therefore, electrons captured already in the deep interface states cannot escape during falling $V_{TG}$, while electrons with relatively small $\tau_e$ near the band edge can escape from the interface states. This is the region (ii). Therefore, the difference in the area integral of $I_{S/D}$ for the ramping and falling of $V_{TG}$ provides the charge $\Delta Q$ for electrons captured to $D_{it}$ in the region (iii). When $V_{init}$ is changed to $V_{init}$' and $V_{fin}$ is fixed, $\Delta Q$ is also changed to $\Delta Q$'. In this case, the average $D_{it}$ in the voltage range of $V_{init}$' - $V_{init}$ can be defined as $(\Delta Q - \Delta Q') / (e^2 \times (V_{init}' - V_{init}))$, where $e$ is elementary charge. It is assumed that the Fermi energy has a linear relation with $V_{TG}$ for simplicity.

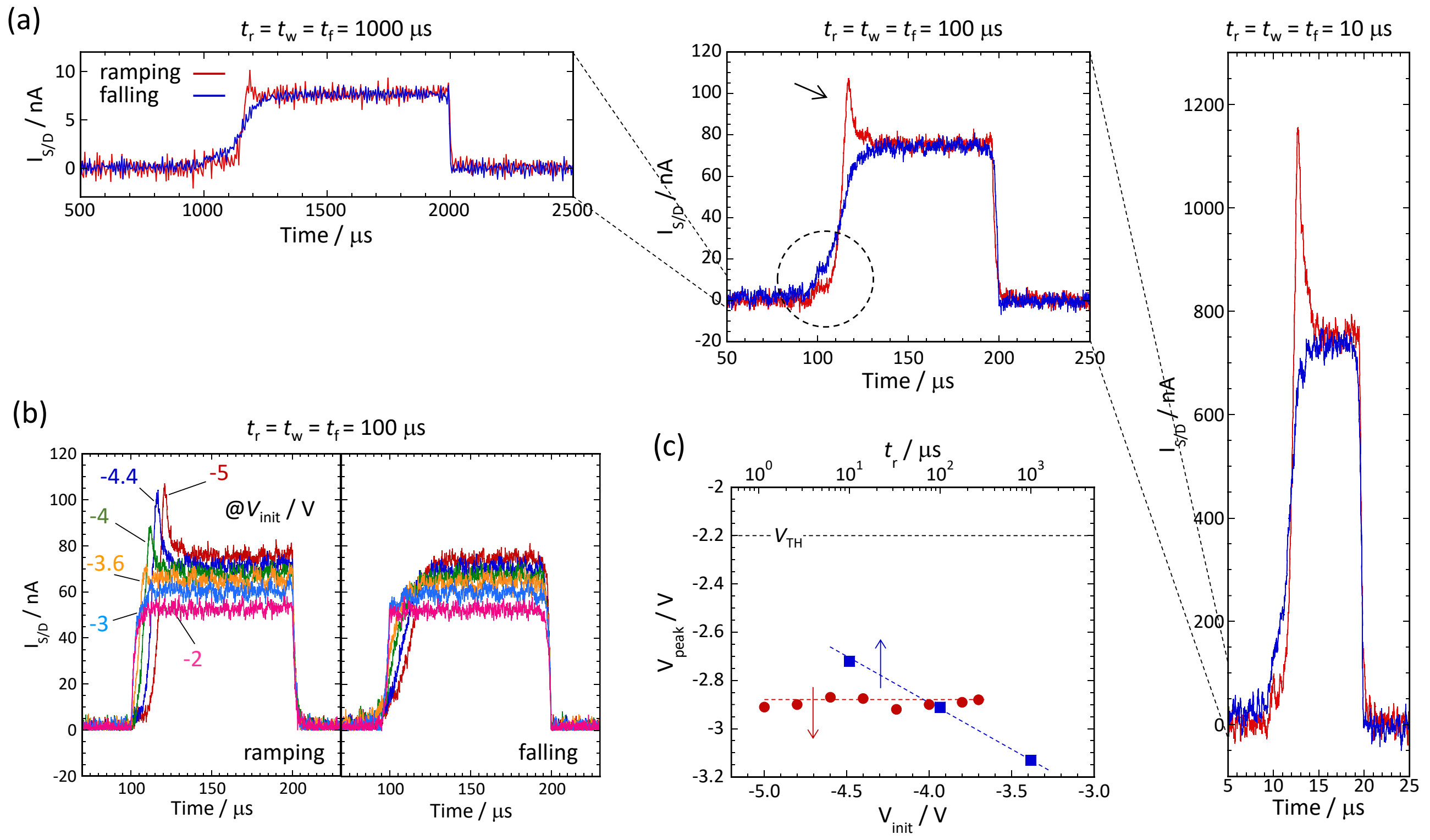

**Figure 3.** (a) $I_{S/D}$ as a function of time at different ramping/waiting/falling times. $V_{init}$ and $V_{fin}$ are fixed at -5 V and 5 V, respectively. The transient current curves for the falling stage are inverted for comparison with the transient curves for the ramping stage. (b) $I_{S/D}$ as a function of time at $t_r = t_w = t_f = 100$ µs for ramping (left) and falling (right). $V_{init}$ is changed from -5 V to -2 V, while $V_{fin}$ is fixed at 5 V. The transient current curves for the falling stage are inverted for comparison with the transient curves for the ramping stage. (c) $V_{peak}$ as a function of $V_{init}$ and $t_r$ (= $t_w = t_f$).

Before quantitatively estimating $D_{it}$, one important caution should be taken into consideration. At $V_{init}$, the system should be in equilibrium, that is, the interface states above $E_F$ at $V_{init}$ should be open as shown in **Figure 2(b-ii)**. However, $MoS_2$ FET is normally on-state at $V_{TG}$ = 0 V. As shown in **Figure 2(b)**, when $V_{TG}$ is set to $V_{init}$ from 0 V, the system may not be in equilibrium, because electrons must escape from the interface states. Therefore, a proper $V_{init}$ keep time, which is indicated by $t_k$ in **Figure 2(b)**, should be found. **Figure 4(b)** shows $\Delta Q$ as a function of $t_k$ at different $V_{init}$. As expected, $\Delta Q$ increases with increasing $t_k$ and saturates for $V_{init}$ = ~2.5 - 4 V because $Q$ for region (i) increases, and $Q$ for region (ii) is almost constant. On the other hand, $\Delta Q$ continuously increases for $V_{init}$ = -5 V due to the interface states newly generated by the electrical stress. Therefore, $V_{init}$ > -3.5 V and $t_k$' = 10 s are adapted as the measurement condition for $D_{it}$. **Figure 4(c)** shows $\Delta Q$ as a function of $V_{init}$. Two different regions are evident, and the transition voltage is close to $V_{TH}$. Since the electron capture has already been finished before $V_{TH}$, $\Delta Q$ observed at $V_{TG} > V_{TH}$ might result from the same origin as the hysteresis in the *I-V*, as shown in the inset of **Figure 4(c)**. Therefore, the hatched region is examined herein. **Figure 4(d)** shows $D_{it}$ as a function of $V_{TG}$ estimated from the time domain CP. The $D_{it}$ values are roughly constant in the range of $10^{12}$ - $10^{13}$ cm$^{-2}$ eV$^{-1}$. Although the $D_{it}$ peak related to S vacancy is often reported for multilayer $MoS_2$,[3,5] it has not been observed in the present monolayer $MoS_2$, which is consistent with the $D_{it}$ calculated from *S.S.* It should be noted that the current peak for electron capture in **Figure 3(a)** is not related with the S vacancy because $V_{peak}$ shifts with ramping speed in **Figure 3(c)** and the energy level for the S vacancy is much deeper than $V_{TH}$.[28-30] Since the inherently high $MoS_2$ channel resistance prevents the extraction of $D_{it}$ from the frequency dispersion in *C-V* analysis on the $MoS_2$ FET structure, the present time domain CP provides an alternative method.

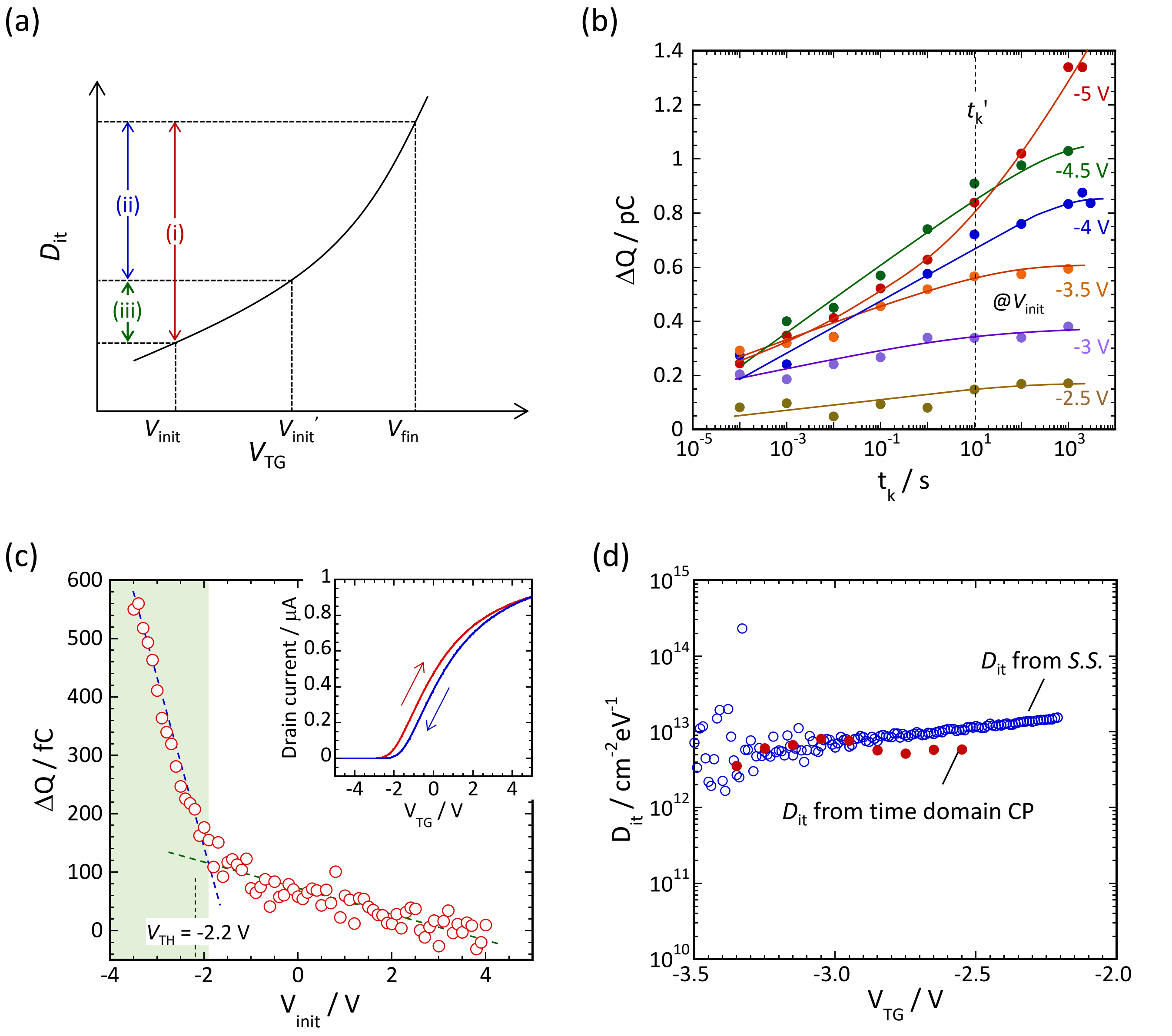

**Figure 4.** (a) Schematic to explain $D_{it}$ extraction. (b) $\Delta Q$ as a function of $t_k$ for different $V_{init}$. (c) $\Delta Q$ as a function of $V_{init}$. The inset shows $I_D$ - $V_{TG}$ for round sweep measurement. (d) $D_{it}$ as a function of $V_{TG}$ estimated from time domain CP. For comparison, $D_{it}$ calculated from *S.S.* is also plotted.

In summary, time domain CP analysis on monolayer $MoS_2$ FET elucidated the elementary process for capture and emission of electrons. The electrons induced during the top gate voltage ramping are suddenly trapped to the interface states, resulting in the peak behavior. This occurs before the channel is completely formed. On the other hand, the emission process is much slower than the electron capture process, resulting in the broad tail behavior. Based on the differences between $\tau_c$ and $\tau_e$, $D_{it}$ is evaluated to be in the range of $10^{12}$ - $10^{13}$ cm$^{-2}$ eV$^{-1}$, which is comparable with that estimated from *S.S.* The present time domain CP can be used as an alternative method for $D_{it}$ estimation.

**Supplementary material**

See the supplementary material for the time domain charge pumping data for Si-MOSFET.

**Acknowledgements**

This research was supported by the JSPS A3 Foresight program, Core-to-Core Program, A. Advanced Research Networks, JSPS KAKENHI Grant Numbers JP25107004, JP16H04343, JP16K14446, and JP26886003, Japan.

Supplementary information

# Direct observation of electron capture & emission processes by the time domain charge pumping measurement of $MoS_2$ FET

**Koki Taniguchi,[1] Nan Fang[1] and Kosuke Nagashio[1,*]**

[1]Department of Materials Engineering, The University of Tokyo, Tokyo 113-8656, Japan

**E-mail:** *nagashio@material.t.u-tokyo.ac.jp

Corresponding author, *nagashio@material.t.u-tokyo.ac.jp

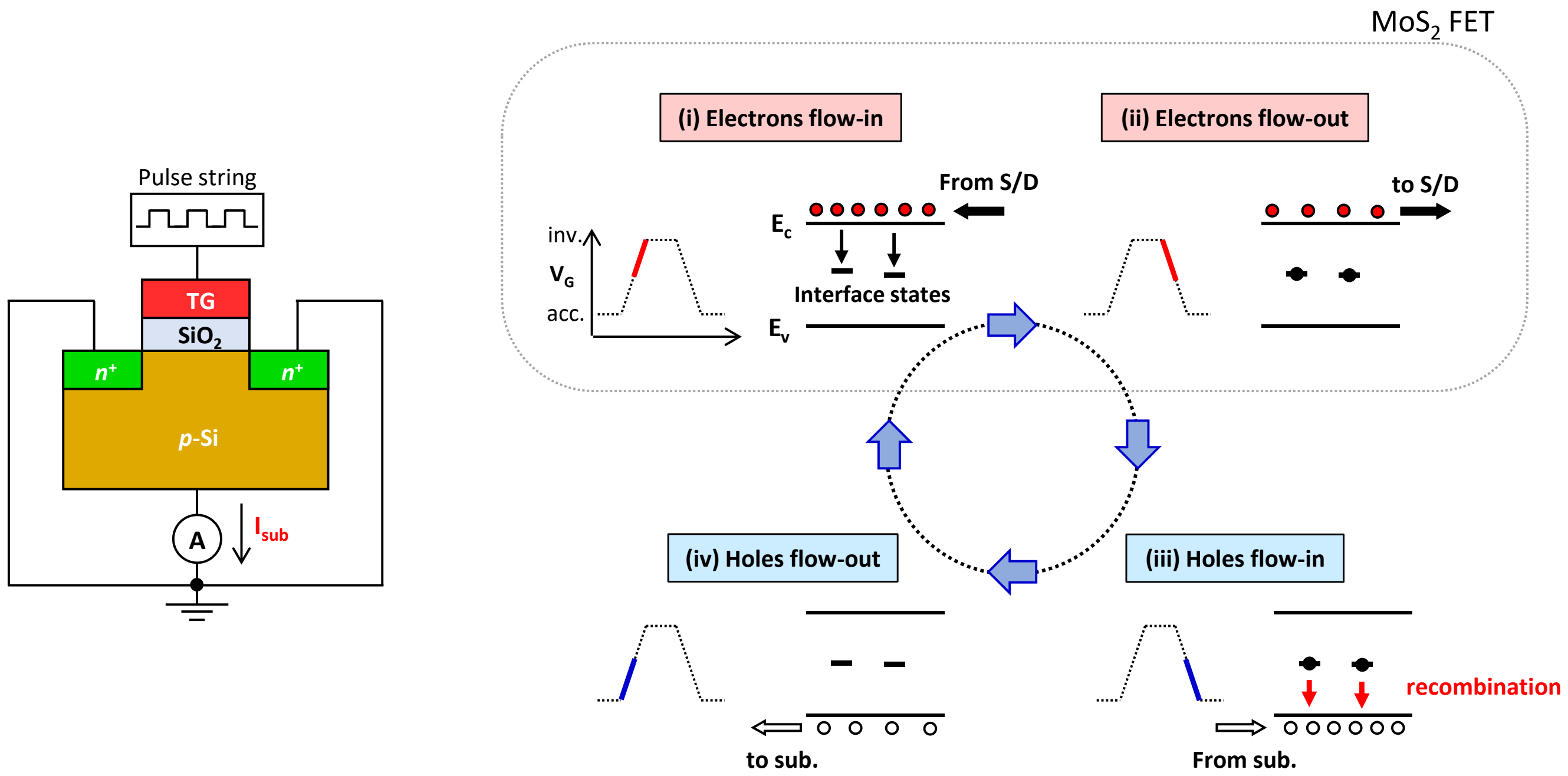

**Figure S1**: Schematic of Si-N-MOSFET and the measurement set up for conventional CP. There are four processes in the one pulse of $V_G$ between the accumulation and the inversion. (i) electrons flow-in, where the electrons are captured in interface states. It should be noted that electrons are also captured during the "high state" of the gate pulse (constant voltage region), depending on the time constant of the traps. (ii) electrons flow-out, where some of electrons captured in the interface states cannot flow out. Therefore, the number of electrons which flow out to S/D is reduced. (iii) holes flow-in, where the recombination of captured electrons and free holes takes place. Again, free holes also capture the electrons during the "low state" of the gate pulse. (iv) holes flow-out, where the number of holes which flow out to the substrate is reduced. In this single CP process, the current flows from the substrate to the S/D through the interface

states. Experimentally, by applying the $V_G$ pulse string using the pulse generator, this substrate current (CP current) is monitored by the DC ammeter. Therefore, the conventional CP does not require the fast acquisition set up. However, this conventional CP is not applicable to $MoS_2$ FET, because source/drain electrodes are connected only to the conduction band due to the Fermi level pinning. At this moment, it is difficult to connect to the valence band. Therefore, the recombination current cannot be measured. Instead of that, the time domain CP measurement is possible only for the (i) and (ii) processes, as shown by the dashed rectangular because the elemental process of electron flow-in (capture) and electron flow-out (emission) can be extracted by fast measurement set up.

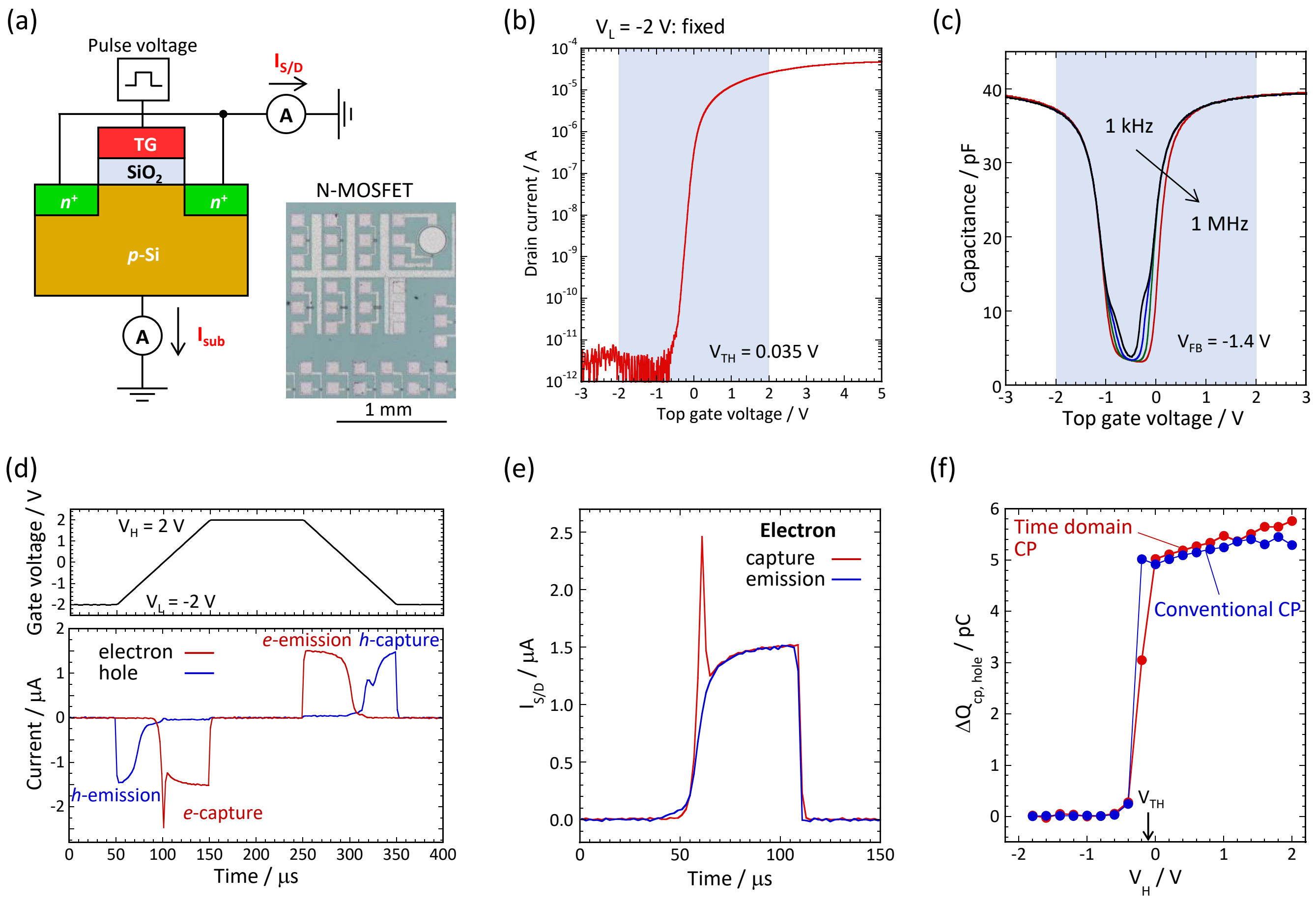

**Figure S2**: (a) Schematic of Si N-MOSFET and measurement set up. Optical micrograph of Si N-MOSFET. All the measurement was carried out at the room temperature. (b) $I_D$-$V_D$ transfer curve at $V_D$ = 0.1 V. The threshold voltage is $V_{TH}$ = 0.035 V. (c) *C*-*V* curves at different frequency. The flat band voltage is $V_{FB}$ = -1.4 V. (d) Time domain CP measurement for Si N-MOSFET. Top: Top gate voltage pulse as a function of time. Bottom: Electron current detected through the source/drain (shown by red curve) and hole current detected through the substrate (shown by blue curve). (e) Comparison of electron current for capture and emission, where electron capture current is "inverted" to compare two data. (f) $\Delta Q_{CP,hole}$ as a function of $V_H$. $\Delta Q_{CP,hole}$ is calculated from the difference in the time-integral of $I_{S/D}$ (the area of (e)) for capture and emission.